\documentclass[aip,jmp,
preprint,%
]{revtex4-1}

\usepackage{dcolumn}
\usepackage{bm}

\input epsf
\usepackage{amssymb}
\usepackage[dvips]{graphicx}
\usepackage{amscd}
\usepackage{mathrsfs}
\usepackage{amsmath}
\usepackage{cancel}

\newcommand{\be}{\begin{equation}}
\newcommand{\ee}{\end{equation}}
\newcommand{\beqa}{\begin{eqnarray}}
\newcommand{\eeqa}{\end{eqnarray}}
\newcommand{\nn}{\nonumber}

\begin{document}

\preprint{AIP/123-QED}

\title[The velocity operator in quantum mechanics in noncommutative space]{The velocity operator in quantum mechanics in noncommutative space}%

\author{Samuel Kov\'a\v cik}
\author{Peter Pre\v snajder}%
 \email{presnajder@fmph.uniba.sk}
\affiliation{
Faculty of Mathematics, Physics and Informatics, Comenius University of Bratislava, Mlynsk\'a dolina F2, Bratislava, Slovakia
}%

\date{\today}

\begin{abstract}
We tested  the consequences
of a noncommutative (NC from now on) coordinates $x_k$, $k = 1,2,3$ in the framework of quantum mechanics. We restricted ourselves to 3D rotationally invariant NC
configuration spaces with dynamics specified by the Hamiltonian $\hat{H} = \hat{H}_0 + \hat{U}$, where $\hat{H}_0$ is an analogue of kinetic
energy and $\hat{U} = \hat{U}(\hat{r})$ denotes an arbitrary rotationally invariant potential. We introduced the velocity operator by
$\hat{V}_k = - i [\hat{X}_k, \hat{H}]$ ($\hat{X}_k$ being the position operator), which is a NC generalization of the usual gradient operator
(multiplied by $-i$). We found that the NC velocity operators possess various general, independent of potential, properties: $1)$ uncertainty
relations $[\hat{V}_i,\hat{X}_j]$ indicate an existence of a natural kinetic energy cut-off, $2)$ commutation relations
$[\hat{V}_i,\hat{V}_j] = 0 $, which is non-trivial in the NC case, $3)$ relation between $\hat{V}^2$ and $\hat{H}_0$ that indicates the existence of
maximal velocity and confirms the kinetic energy cut-off, $4)$ all these results sum up in canonical (general, not depending on a particular form of the central potential) commutation relations of Euclidean group $E(4) = SO(4)\triangleright T(4)$, $5)$  Heisenberg equation for the velocity operator, relating acceleration  $\dot{\hat{V}}_k = -i[\hat{V}_k, \hat{H}]$ to derivatives of the potential.
\end{abstract}

\maketitle

\section{\label{intro}Introduction}
\subsection*{Motivation}

The idea that the mathematical continuum is not an adequate model for physical space or space-time was suggested by W. Heisenberg long time ago in connection with UV divergencies that appeared in standard quantum field theory. In 1947 were published papers by H. S. Snyder \cite{Snyder} and C. N. Yang \cite{Yang} devoted to quantized space-time: The coordinates were noncommuting, thus preventing the full localization of space-time points. Roughly at the same time J. A. Wheeler suggested that the space-time should be quantized in order to formulate quantum theory of Einstein gravity, see \cite{Wheeler}. This direction has not been developed much further, mainly due to the success of the renormalization theory approach to quantum field theory.

Approximately four decades later the idea of quantum space was pushed forward when the idea of noncommutative geometry was introduced in \cite{Con}, and in the form of matrix geometry it was formulated in \cite{Mad1}. The need for non commutative geometry has been motivated in \cite{DFR} by investigating combination of quantum physics and gravity. The basic argument goes as follows: If we keep shortening the wavelength of a photon $\lambda_\gamma$, its radius of the event horizon keeps growing larger and larger ($r_{S\gamma} = 2 \kappa h/ \lambda_\gamma c^3$), until eventually those two become equal and the photon is hidden under its event horizon. This happens for $\lambda_\gamma \propto l_{Planck}$ - so we cannot distinguish two points if their distance is smaller than the Planck's length. This led to a set of specific uncertainty relations for space-time coordinates $x_\mu$, $\mu = 0, 1, 2, 3$, that are fulfilled provided the coordinates satisfy commutation relations
\begin{equation}
[x_\mu,\,x_\nu]\ =\  i\, \theta_{\mu\nu}\,,\ \ \theta_{\mu\nu}\ -\ \mbox{constant}\,.
\end{equation}
However, such theories violate Lorentz invariance. Later it was shown that field theories in such noncommutative space-time emerge as effective low-energy limits of string theories \cite{string}.

It was a challenge to ask if we could translate physics into a model of space, whose close points cannot be exactly localized (on some scale, which is described by $\lambda$ - not to be confused with any wavelength) and even more interesting to find, that this is in fact possible (we can obtain results which reproduce well known results in the limit of $\lambda = 0$, what corresponds to the ordinary space). The use of non commutative geometry presents a hope to remove UV divergences from quantum field theory and, eventually, to quantize gravity. Both mentioned concepts are extremely complicated, with partial successes only and various unsolved problems.

Therefore, it would be interesting to reverse the problem, not to try to formulate a new fundamental theory, but to test the consequences of noncommutative (NC) configuration space in the framework of a well-defined quantum mechanics. Various classical QM problems (Aharonov-Bohm effect, hydrogen atom, ...) has been investigated in noncommutative configuration spaces, however, with the noncommuativity violating rotational invariance, see \cite{AB}, \cite{CP}. A spherical well in rotationally invariant 2D NC space has been described in \cite{Scholtz}.

The main goal of the paper is to investigate QM in 3D NC space described by coordinates satisfying rotationally invariant commutation relations
\begin{equation} \label{NC relation}
[x_i,\,x_j]\ =\ 2 i\, \lambda\, \varepsilon^{ijk}\, x_k ,\ \ \ i,j,k\,=\,1,2,3,
\end{equation}
where $\lambda$ describes the scale of the noncommutativity of the space ($\lambda $ is not fixed within our model). This commutation relations define NC configuration space that is a noncommutative analog of the usual 3D configuration space with the origin eliminated: ${\bf R}^3_0\,=\,{\bf R}^3\setminus \{0\}$. Such NC space was introduced in \cite{GP1} and \cite{GP2}, and applied to the investigation of NC Coulomb problem. The model is exactly solvable, possessing various expected and some unexpected features.

Here we shall extend this approach to the case of a particle moving in NC space in an arbitrary central potential:  $\hat{H}=\hat{H}_0 + \hat{U}(\hat{r})$, where $\hat{H}_0$ is NC analogue of the kinetic term introduced in \cite{GP1},\cite{GP2} and $\hat{U}=\hat{U}(\hat{r})$ is a given central potential. Besides the Hamiltonian, we shall introduce, in terms of the NC coordinates, the position operator $\hat{X}_j$, the angular momentum operator $\hat{L}_j$ and the velocity operator defined as
%
\begin{equation}\label{NC velo}
\hat{V}_j\ =\ -i\, [ \hat{X}_j ,\, \hat{H}]\ =\ -i\, [\hat{X}_j,\, \hat{H}_0]\,.
\end{equation}
This velocity operator is self-adjoint on a proper domain. We shall not consider NC analog of the momentum operator $\hat{P}_j$ corresponding to the shift in configuration space, since for spaces with nontrivial topology or boundary, such as ${\bf R}^3_0$ is, the definition of self-adjoint shift operators is not straightforward and requires special attention.

For NC states $\psi$ that possess classical analogue $\psi(x_j)$ in the commutative limit $\lambda \,\to\,0$ we have obtained following general relations among basic observables, not depending on a specific form of $\hat{U}(\hat{r})$:
\be \label{res I} [\hat{L}_i,\, \hat{L}_j] \,= i\, \varepsilon_{ijk}\, \hat{L}_k, \ \ \
[\hat{L}_i,\, \hat{X}_j] \,=\, i\, \varepsilon_{ijk}\, \hat{X}_k, \ \ \
 [\hat{X}_i,\, \hat{X}_j] \,=\, 2i\,\lambda^2\,  \varepsilon_{ijk}\, \hat{L}_k\,.
\ee
\be \label{res II} [\hat{L}_i,\, \hat{V}_j] \,=\, i\, \varepsilon_{ijk}\, \hat{V}_k, \ \ \ [\hat{V}_i,\, \hat{X}_j] ,=\, -i\,\delta_{ij}\, \left( 1-\lambda^2 \hat{H}_0 \right), \ \ \ [\hat{V}_i,\, \hat{V}_j] \,=\, 0\,.
\ee
\be \label{res III} \hat{V}_k^2 \,=\, 2\,\hat{H}_0 \,-\,\lambda^2\, \hat{H}^2_0, \ \ \
\hat{L}_k\,\hat{V}_k \,=\, 0\, .
\ee
\begin{equation}
\label{res IV} {\hat{A}}_j \,=\,  -i\, [\hat{V}_j,\, \hat{H}] =  -i\, [\hat{V}_j,\, \hat{U}(\hat{r})] \,.
\end{equation}
The first line tells us that components of the orbital momentum operator and the position operator generate $SO(4)$ kinematical symmetry; introducing properly $V_4$, equations (\ref{res II}) extend kinematical symmetry to $E(4)$ (the group of isometries of 4D Euclidean space); the quadratic relations (\ref{res III}) specify the $E(4)$ representation in question to the scalar one.  Last equation  represents the Heisenberg equation of motion for the acceleration $\hat{A}_j$ in terms of NC derivatives of $\hat{U}(\hat{r})$, the commutator is evaluated below.

We would like to point out that these formulas can be extended to magnetic monopoles. This problem is under scrutiny.

The paper is organized as follows. In Section 2 we briefly describe the formulation of NC QM, Section 3 contains the derivation of all mentioned results. The last Section 4 is devoted to conclusions, some technicalities are postponed to the Appendices.

\section{QM in NC space}

\subsection*{Construction of a NC space}

The first thing to begin with is a construction of NC space. From the ordinary QM we know, that the impossibility of measuring exact position and momentum of a particle originates in the Heisenberg's uncertainty principle $\Delta _x \Delta _p \geq \frac{\hbar}{2}$, which follows from the relation $[\hat{x}, \hat{p}]=i \hbar$. With this in mind we expect some kind of commutation relation to describe the coordinates of a fuzzy space.

The first choice one can think of is $[x^i, x^j] = i \theta^{ij}$, where $\theta^{ij}$ is \textit{some} antisymmetric matrix. This choice however lacks rotational invariance, what may lead to severe complications of computations.

Instead, we shall follow the construction proposed in \cite{GP1} and \cite{GP2}. We postulate the rotationally invariant relation \eqref{NC relation}
and realize it with two sets of bosonic creation (c) and annihilation (a) operators (from now on only 'c/a operators') satisfying well known relations
\begin{equation} \label{bosonic}
[a_\alpha ,\, a^+_\beta]\ =\ \delta_{\alpha \beta} ,\ \ \ [a_\alpha ,\, a_\beta]\ =\ [a^+_\alpha,\, a^+_\beta]\ =\ 0,\ \ \  \alpha, \beta = 1,2 \,.
\end{equation}
They act in an auxiliary Fock space ${\cal F}$ spanned by normalized vectors $| n_1, n_2\rangle$
\begin{equation}
|\, n_1, n_2 \rangle\ =\ \frac{(a_1^+)^{n_1}(a_2^+)^{n_2}}{\sqrt{n_1!n_2!}}\ |\, 0,0 \rangle \,.
\end{equation}

One can easily check that the spatial coordinates defined as
\begin{equation}\label{X}
x_j\ =\ \lambda a^+ \sigma_j a\ =\ \lambda \sigma ^j _{\alpha \beta}a^+_\alpha a_\beta,\ \ \ j=1,2,3\,,
\end{equation}
satisfy \eqref{NC relation} ($\sigma_j$ are the Pauli matrices). In addition we introduce radial coordinate defined as
\begin{equation} \label{r}
r\ =\ \lambda\, ( N+1)\ =\ \lambda\, (a^+_\alpha a_\alpha +1) \,.
\end{equation}
It is interesting to note, that now $x^2\, \neq\, r^2$, but instead
\begin{equation} \label{r^2}
x^2\ =\ r^2 \,-\, \lambda ^2 \,.
\end{equation}

{\it Note 1}: This choice of operators $x_j$, $j = 1,2,3$, is based on a deformation quantization (fuzzification) of a complex plane $\mathcal{C}^2$ endowed with a flat Poisson bracket
\be \label{z}  \{ z_\alpha, \bar{z}_\beta \}\ =\ - i\,\delta_{\alpha \beta },\ \ \ \{ z_\alpha, z_\beta \}\ =\ \{ \bar{z}_\alpha, \bar{z}_\beta \}\ =\ 0\,,
\ee
where ($z_1,z_2) \in \mathcal{C}^2$ and the bar denotes complex conjugation. Putting
\begin{equation} \label{sympl}
\xi_j\ =\ z^+ \sigma ^j z,\ \ \ \Rightarrow \ \ \ \{ \xi_i, \xi_j \}\ =\  2 \, \varepsilon_{ijk}\, \xi_k \,,
 \end{equation}
we obtain commutative versions of equations \eqref{X} and \eqref{r}. Performing deformation quantization of \eqref{sympl} we obtain noncommutative algebra of functions $A = A(\xi_j)$, $\xi = (\xi_1,\xi_2,\xi_3)$, endowed with a star-product which is isomorphic to the algebra of operators $ A = A(x)$, $x = (x_1,x_2,x_3)$, ($x_k$ are operators in Fock space given in \eqref{X}). We shall use operator realization of NC, since it is better adapted for our purposes (however, star-product may be simpler in particular cases, such as evaluating of commutative limits).

\subsection*{Construction of the Hilbert space $\mathcal{H}$ and some important operators}

The Hilbert space ${\cal H}_\lambda $ is a completion of the linear space of operators in the Fock space spanned by the monomials of the form
\begin{equation} \label{states}
 (a_1^+)^{m_1}(a_2^+)^{m_2}(a_1)^{n_1}(a_2)^{n_2} ,\ \ \ m_1 +m_2 - n_1-n_2\, =\, \kappa\, =\, 0 \,,
\end{equation}
with respect to the norm defined as
\begin{equation} \label{ss}
||\psi||^2 = 4\pi \lambda ^2 Tr [ \psi^+ \,r\, \psi] \,.
\end{equation}
The constants are chosen so that the  volume of a ball with radius $r\,=\,\lambda (N+1)$ is $V_r\,=\, \frac{4}{3}\pi r^3 + O(\frac{\lambda}{r})$, i.e., for $r \gg \lambda $ the volume approaches its standard value, see \cite{GP1} and \cite{GP2}.

In this paper we study only states for which $\kappa = 0$, that according to \eqref{X} can be expressed as $\psi=\psi(\vec{x})$. A slight generalization $m_1 +m_2 - n_1-n_2 = \kappa \neq 0$ leads to results known from theories of magnetic monopoles. 

The angular momentum operators $\hat{L}_k$ and the position operators $\hat{X}_k$, $k = 1,2,3$, are defined as
\beqa
\nn \hat{L}_k \psi &=& \frac{1}{2\lambda}\,[x_k, \psi]\ = \ \frac{1}{2\lambda}\,(x_k \,\psi \,-\,\psi \, x_k  )\,, \\
\label{LX} \hat{X}_k \psi &=& \frac{1}{2}\,(x_k \,\psi \,+\,\psi \, x_k  )\,,\ \ \ k\,=\,1, 2, 3\,.
\eeqa
The angular momentum and coordinate operators can be given as left and right multiplications by NC coordinates
\begin{equation}
\hat{X}^L_k \,\psi\ =\ x_k\, \psi ,\ \ \  \hat{X}^R_k\, \psi\ =\ \psi\, x_k \,.
\end{equation}

It can be easily verified that angular momentum and position operators satisfy commutation relation \eqref{res I}. Introducing
notation
\begin{equation} \label{L-SO4}
\hat{L}_{ij}\ =\ \varepsilon_{ijk}\,\hat{L}_k,\ \ \ \hat{L}_{k4}\ =\ -\,\hat{L}_{4k}\ =\ \frac{1}{\lambda}\, \hat{X}_k\,,
\end{equation}
commutation relation \eqref{res I} take explicit $SO(4)$ invariant form
\begin{equation}
[\hat{L}_{ab},\hat{L}_{cd}]\ =\ i\,(\delta_{ac}\,\hat{L}_{bd}\,-\, \delta_{bd}\,\hat{L}_{ac} )\,.
\end{equation}

It can be easily verified that $\hat{X}_k$, $k = 1,2,3$, transform as a 3D vector (with respect to rotations generated by $\hat{L}_k$, $k = 1,2,3$). Similarly, the radial coordinate operator $\hat{r}$ defined as
\begin{equation}
\hat{r}\, \psi\ \equiv\ r\, \psi\ =\ \psi\,r \,.
\end{equation}
is an $SO(4)$ scalar $[\hat{L}_{ab},\hat{r}]\, =\, 0$ (because $[x^k,r]=0$ as follows from \eqref{bosonic}, \eqref{r} and \eqref{LX}).

Operators $\hat{L}_k$ act on states
\begin{equation} \label{states2}
\psi_{jm}\ =\ \sum \limits _{(jm)} \frac{(a_1^+)^{m_1}(a_2^+)^{m_2}}{m_1!m_2!}\,\mathcal{R}_j(r)\,\frac{(a_1)^{n_1}(-a_2)^{n_2}}{n_1!n_2!}\,,
\end{equation}
where the summations goes over all non negative integers $m_1, m_2, n_1, n_2$ satisfying restrictions
\begin{equation}
m_1+m_2+n_1+n_2\ =\ 2j,\ \ \ \ m_1-m_2-n_1+n_2\ =\ 2m \,,
\end{equation}
in a standard way
\begin{equation}
\hat{L}_k^2\,\psi _{(jm)}\ =\ j(j+1)\, \psi _{(jm)},\ \ \ \hat{L}_3\,\psi_{(jm)}\ =\ m\,  \psi _{(jm)}\,.
\end{equation}

Another important operator to define is the free Hamiltonian, or the Laplace operator (what is the same up to a multiplicative constant). Let us present a line of thoughts which leads to it. First step: the Laplace operator in ordinary space is a second order differential operator, so in NC space we expect a double commutator ($\hat{\Delta}_{\lambda} \psi \propto [.,[.,\psi]]$). Second step: we choose for the Laplace operator the simplest rotationally invariant double commutator $\hat{\Delta}_{\lambda} \psi \propto [a^+_\alpha,[a_\alpha,\psi]]$ (note, that due to the Jacobi identity the order of c/a operators is arbitrary). Next we require $\hat{\Delta}_{\lambda}$ being hermitian with respect to \eqref{ss}, what leads to $\hat{\Delta}_{\lambda} \psi \propto \frac{1}{r} [a^+_\alpha,[a_\alpha,\psi]]$. Still we have to fix the numerical factor. To do so we examine its action on some simple function, i.e., we calculate $\hat{\Delta}_\lambda R(r)$. The final result is (see \cite{GP1} and \cite{GP2}):
\begin{equation} \label{NCDelta}
\hat{\Delta}_\lambda\, \psi\ =\ - \frac{1}{\lambda r}\, [a^+_\alpha ,\, [a_\alpha ,\, \psi]]\,,
\end{equation}
or equivalently
\begin{equation}
\hat{H}_0\, \psi\ =\ \frac{1}{2m \lambda r}\, [a^+_\alpha ,\, [a_\alpha ,\, \psi]]\,.
\end{equation}

The Hamiltonian $\hat{H}_0$ supplemented by the Coulomb potential $\hat{U}\, =\, - \frac{q}{\hat{r}} $ has been used in \cite{GP1} and \cite{GP2} to study the NC hydrogen atom problem. Besides solutions analytic in $\lambda $ that reduce to the standard expressions in the commutative limit $\lambda\, \to\, 0$,  solutions singular in $\lambda $ corresponding to energies $E\,\sim\,\lambda^2$ that disappear in the commutative limit also have been found.

{\it Note 2}: The construction of NC version of the Laplace operator may seem a little bit ad hoc, but we shall show that in the commutative limit the operator $\hat{\Delta}_\lambda$ reduces to the standard Laplace operator. Let us consider commutative version of equation \eqref{NCDelta} in terms of the Poisson bracket \eqref{z}:
\begin{equation} \label{CDelta}
\hat{\Delta}_0\, \psi(\xi_j)\ =\ \frac{1}{r}\, \{\bar{z}_\alpha ,\, \{z_\alpha ,\, \psi(\xi_j)\}\}\ =\ \frac{1}{r}\,\partial_{z_\alpha} \partial_{\bar{z}_\alpha}\,\psi(\xi_j)\,,
\end{equation}
where $\xi_k\ =\ z^+ \sigma_k z$ and $r = z^+ z$. Taking  the chain rule for derivatives and properties of the Pauli matrices into account it follows directly that
\begin{equation}
\hat{\Delta}_0\, \psi(\xi_j)\ =\ \partial_{\xi_k} \partial_{\xi_k}\, \psi(\xi_j)\ =\ \Delta\,\psi(\xi_j)\,.
\end{equation}
Thus $\hat{\Delta}_\lambda$ is a $\lambda$-deformation of the usual Laplace operator.

\section{The velocity operator}

\subsection*{Definition}
Let us define the velocity operator by the Heisenberg equation as the commutator of the coordinate operator with the Hamiltonian $\hat{H}=\hat{H}_0 + \hat{U}(r)$ as follows:
\begin{equation} \label{V}
\hat{V}_j \, \psi \,=\,  -i\, [ \hat{X}_j,\, \hat{H}]\,\psi \,=\,
-i\, [ \hat{X}_j,\,  \hat{H}_0]\,\psi \,,
\end{equation}
where we have taken into account that $\hat{X}_j$ commutes with  $\hat{U}=\hat{U}(\hat{r})$ as a consequence of $[x^j, r]=0$. Thus for all radial potentials $\hat{U}=\hat{U}(\hat{r})$ the velocity operator is determined only by the commutator with the kinetic part of the Hamiltonian which is equal to
\begin{equation}
\label{V1} \hat{V}_j\,\psi\ =\ -\frac{i}{2r}\,\sigma ^j _{\alpha \beta}\, \left(a^+_\alpha\, [a_\beta ,\, \psi]\, -\, a_\beta\, [ a^+_\alpha ,\,\psi] \right) \,.
\end{equation}

From its construction, being defined by the commutator of hermitian operators $\hat{X}^i$ and $\hat{H}_0$ (see \ref{V}), it is obvious that the velocity operator is hermitian as well.

We shall now investigate its action on some basic objects to prove that it really behaves as the NC generalization of the gradient operator, as was proposed earlier.

The calculation of $\hat{V}_i x_j$ will be the only complete calculation outside of the Appendix, because it should give the reader the basic idea about the calculations within our model of NC QM, and is short enough not to distract them. Inserting $x_j \,=\,\lambda \sigma^j_{\gamma \delta} \, a^+_\gamma a_\delta $ into \eqref{V1} we obtain
\beqa
\nn \hat{V}_i \,x_j &=& -\frac{i\lambda}{2r}\,\sigma^i_{\alpha \beta}\, \sigma^j_{\gamma \delta}\,\left(a^+_\alpha\, [ a_\beta ,\, \lambda\,  a^+_\gamma a_\delta]\, -\, a_\beta\, [ a^+_\alpha ,\, \lambda\, a^+_\gamma\, a_\delta] \right) \\
\nn &=& - \frac{i \lambda}{2r}\, \sigma^i_{\alpha \beta}\, \sigma^j_{\gamma \delta}\, \left(a^+_\alpha\, a_\delta\, \delta_{\beta \gamma}\, +\, a_\beta a^+_\gamma \,\delta_{\alpha \delta} \right)
\eeqa
now using \eqref{bosonic} in form $a_\beta a^+_\gamma\, =\, a^+_\gamma a_\beta\, +\, \delta_{\beta \gamma}$ we get
\begin{equation} \label{VX}
\hat{V}_i \,x_j\ =\ -\frac{i\lambda}{2r}\, \left[a^+_\alpha a_\beta\, (\sigma^i \sigma^j\,+\,\sigma^j \sigma^i)_{\alpha \beta})\, +\, Tr(\sigma^j \sigma^j) \right]\ =\ -i\, \delta^{ij} \,.
\end{equation}
In the last step we have used formulas $\{\sigma^i, \sigma^j\}_{\alpha \beta}\, =\, 2 \delta^{ij}\delta_{\alpha \beta}$, $r\,=\, \lambda(a^+_\alpha a_\alpha+1)$ and the trace relation $Tr(\sigma^i \sigma^j)\, =\, 2\delta^{ij}$.
Similarly, one can verify that
\begin{equation} \label{Vr}
\hat{V}_j\, f(r)\ =\ -i\, \frac{x_j}{r}\, f_\lambda'(r)\,,
\end{equation}
where
\begin{equation} \label{f'}
 f_\lambda'(r)\ =\ \frac{f(r+\lambda)-f(r-\lambda)}{2\lambda}\ .
\end{equation}

So, $i \hat{V}_i$ resembles the gradient operator as much as  possible in the NC space. However, there is one crucial difference - it obeys the Leibniz rule with a correction:
\begin{equation} \label{Leibnitz}
\hat{V}_j\,(AB)\ =\ (\hat{V}_j A)B\,+\,A(\hat{V}_j B)\,+\,{\cal K}_j(A,B)\,,
\end{equation}
where ${\cal K}^j (.,.)$ denotes the correction term
\begin{equation} \label{kor}
{\cal K}^i(A,B)\ = -\, \frac{i}{2r}\,\sigma ^i _{\alpha \beta}\,\left([a^+ _\alpha ,A][a_\beta ,B]-[a_\beta ,A][a^+ _\alpha ,B]\right)\,.
\end{equation}
This correction, as can be seen in \eqref{UR}, vanishes in the commutative limit $\lambda \rightarrow 0$.

{\it Note 3}: The velocity operator in standard QM in terms of complex variables $(z_1,z_2)\,\in\,\mathcal{C}^2$ is given as
\begin{equation}\label{CV}
\hat{V}_k\,\psi\ =\ \frac{i}{2}\,[\xi_k,\hat{\Delta}_0 ]\,\psi\,,\ \ \ \xi_k\,=\,z^+\sigma_k z \,,
\end{equation}
where $\hat{\Delta}_0$ is the usual Laplace operator defined in terms of a double Poisson bracket, see \eqref{CDelta}. Simple calculations give a commutative analogue of \eqref{V1}:
\begin{equation}\label{CV1}
\hat{V}_j\,\psi\ =\ \frac{1}{2r}\,\sigma ^j _{\alpha \beta}\, \left(\bar{z}_\alpha\, \{z_\beta ,\, \psi\}\, -\, z_\beta\, \{ \bar{z}_\alpha ,\,\psi\} \right)\,.
\end{equation}
Acting in \eqref{CV1} on a function of the form $\psi\,=\,\psi (\xi)$ recovers the standard QM result: $\hat{V}_j\,\psi(\xi_k) \,=\, -i\,\partial_{\xi_j} \psi(\xi_k)$.
The velocity operator given in \eqref{CV1}, as a first order linear differential operator with coefficients linear in $z,\,\bar{z}$, obviously satisfies the Leibniz rule. However its $\lambda$-deformation, the NC velocity operator given in \eqref{V1}, contains first order commutators with noncommuting coefficients linear in $a, a^+$, what inevitably leads to the NC corrections to the Leibniz rule.

\subsection*{Kinematic commutators containing  velocity}

After examining the basic properties of the velocity/gradient operator $\hat{V}^j$ we could move on to  something more interesting. Namely, we shall derive commutators \eqref{res II}.

The first commutator tells us that the operators $\hat{V}_j$, $j = 1,2,3$ transform as components of a 3D vector under rotations generated by $\hat{L}_k$, $k = 1,2,3$. Using \eqref{bosonic}, \eqref{LX} and \eqref{V} it follows directly:
\begin{equation} \label{LV}
[\hat{L}_i,\, \hat{V}_k]\ =\ i\, \varepsilon_{ikm}\, \hat{V}_m
\ \ \ \Longleftrightarrow \ \ \ [\hat{L}_{ij},\, \hat{V}_m]\ =\ i\,(\delta_{im}\,\hat{V}_j\,-\,\delta_{jm}\,\hat{V}_i) \,.
\end{equation}

The second commutator is related to  one of the most important relations in the ordinary QM, the Heisenberg's commutation relation $[\hat{p}_i,\hat{x}_j]\,=\,-i \hbar \delta_{ij} $. Now having in mind that we have set $\hbar=m=1$ this can be written as
\begin{equation}
[\hat{V}_i ,\, \hat{X}_j]\ =\ -i\, \delta_{ij}\,.
\end{equation}
In ordinary QM, this can be easily calculated using the Leibniz rule and the fact that $\partial_i x^j = \delta^{ij}$.

Similarly one can proceed in NC QM using \eqref{VX} and \eqref{Leibnitz}. However, the NC correction to the Leibniz rule implies a small, but a crucial, difference. Combining those two equations, we get
\begin{equation} \label{UR}
[\hat{V}_i ,\, \hat{X}_j]\, \psi\ =\ -i\, \delta^{ij}\, \psi\, +\, \frac{1}{2} \left({\cal K}^i(x_j, \psi)\,+\,{\cal K}^i(\psi,x_j)\right) \,.
\end{equation}
By evaluating the correction terms
\begin{eqnarray}
{\cal K}_i (x_j,\psi) &=& -\frac{i}{2r}\,\sigma ^i _{\alpha \beta} \left([a^+_\alpha ,\, x_j]\,[a_\beta ,\, \psi\,]-\, [a_\beta ,\, x_j]\,[a^+_\alpha ,\, \psi] \right)\,, \\ \nonumber
{\cal K}_i (\psi,x_j) &=& -\frac{i}{2r}\,\sigma ^i _{\alpha \beta} \left([a^+_\alpha ,\,\psi]\,[a_\beta ,\, x_j]\,-\, [a_\beta ,\, \psi]\,[a^+_\alpha ,\,  x_j]\right) \,,
\end{eqnarray}
 one finds quite a surprising result, that their sum is equal to $2 i\, \delta^{ij}\, \lambda^2 \, \hat{H}_0\, \psi $. Inserting this into \eqref{UR} yields the result
\begin{equation} \label{uncertainty}
[\hat{V}_i ,\, \hat{X}_j]\ =\ -i\, \delta^{ij} \left(1\, -\, \lambda^2\, \hat{H}_0 \right) \,.
\end{equation}

Let us analyze this result briefly. Firstly, it is exact, as we have not neglected any terms of higher orders in $\lambda$. Moreover, it is obvious that this result presents a small correction , which vanishes for $\lambda=0$. It also defines two important energy scales: $E_0 = 1/\lambda^2$, for which the commutator (\ref{uncertainty}) vanishes, and $E_1 = 2/\lambda^2$, for which the r.h.s of (\ref{uncertainty}) obtains a negative sign, with respect to the ordinary value $-i \delta^{ij}$. Both scales $E_0$ and $E_1$ appeared also in the study of scattering/bound states of NC QM hydrogen atom problem in \cite{GP1} and \cite{GP2}.

%

For the purpose of evaluating the third commutator in \eqref{res II} (and many others later), it is useful to rewrite the velocity operator as
\begin{equation} \label{V2.0}
\hat{V}_j\ =\ \frac{i}{2r}\, \sigma^i_{\alpha \beta}\, \hat{w}_{\alpha \beta}, \ \ \
\hat{w}_{\alpha \beta}\,\psi\ =\ a^+_\alpha\, \psi\, a_\beta\, -\, a_\beta\, \psi\, a^+_\alpha \,.
\end{equation}
Using this notation the commutator $[\hat{V}_i ,\,\hat{V}_j]$ can written as
\begin{equation}
[\hat{V}_i ,\, \hat{V}_j]\ =\ \left(\frac{i}{2} \right)^2 \sigma ^i _{\alpha \beta}\, \sigma ^j_{\gamma \delta}\, \left[ \frac{1}{\hat{r}}\, \hat{w}_{\alpha \beta} ,\, \frac{1}{\hat{r}}\, \hat{w}_{\gamma \delta} \right]\,.
\end{equation}
Using the rule for $[AB,\,CD]$, the commutator in question can be split into 3 separate terms: one of the form $\frac{1}{r^2}[w,w]$ and two of the form $\frac{1}{r}[\frac{1}{r},w]$. We have evaluated them separately (for details see the Appendix):

1) The first term turns out to be equal to $\frac{2i}{\lambda \hat{r}^2}\, \varepsilon^{ijk}\, (\hat{X}^k _L\, -\, \hat{X}^k_R)$, and

2) the sum of other two is equal to $\frac{2i}{\lambda \hat{r}^2}\,\varepsilon^{ijk}\, (-\hat{X}^k_L + \hat{X}^k_R)$.

\noindent Combining those two results, one obtains quite a non-trivial zero
\begin{equation} \label{VV}
[\hat{V}_i,\,\hat{V}_j]\ =\ 0 \,.
\end{equation}
One may ask, if this zero isn't somehow obvious. To answer this question, we investigated this commutator for generalized states $\psi$ containing different number of c/a operators
yielding highly non vanishing, as well as interesting, result.

Remembering that $\hat{X}_j \,=\, \hat{L}_{j4} $ and introducing the 4-th component of the velocity operator by r.h.s. of \eqref{uncertainty} as
\begin{equation} \label{V4}
\hat{V}_4\,\psi\ =\ \left( \frac{1}{\lambda}\,-\,\lambda\, \hat{H}_0\right)\,\psi\ =\ \frac{1}{2r} \left( a^+_\alpha\, \psi\, a_\alpha\, +\, a_\alpha\, \psi\, a^+_\alpha \right)\,,
\end{equation}
we obtain a remarkable result that relations in \eqref{res II} can be written in $SO(4)$ invariant form
\begin{equation} \label{VSO4}
[\hat{L}_{ab},\,\hat{V}_c] \ =\ i\,(\delta_{ac}\,\hat{V}_b\,-\, \delta_{bc}\,\hat{V}_a )\,,\ \ \ [\hat{V}_a,\,\hat{V}_b] \ =\ 0\,,\ \ \ a,b,c\,=\,1,\,...\,,4\,.
\end{equation}
The first commutation relation combines \eqref{LV} and \eqref{uncertainty} and tell us that the velocity operator $\hat{V}_a $, $a\,=\,1,\,...\,,4$, is an $SO(4)$ 4-vector. The second commutation relation extends \eqref{VV} to all four velocity components (the additional relation $[\hat{V}_j,\,\hat{V}_4] \, =\, 0$ follows from the $SO(4)$ invariance).

\subsection*{Quadratic relations for velocity operator} 

First we evaluate relation between the square of the velocity operator $\hat{V}^2_j$ and the free Hamiltonian. This calculation is a bit tricky, for details see the Appendix. One should first evaluate
\begin{equation}
\hat{V}^2_j \ =\ -\frac{1}{4}\ \sigma ^j_{\alpha \beta}\, \sigma^j_{\gamma \delta}\,\frac{1}{\hat{r}}\, \hat{w}_{\alpha \beta}\, \frac{1}{\hat{r}}\,\hat{w}_{\gamma \delta}\,,
\end{equation}
and after that
\begin{equation}
\left(\frac{1}{\lambda}\, -\,\lambda \hat{H}_0 \right)^2 \,.
\end{equation}
Comparing those two expressions one obtains an interesting relation
\begin{equation} \label{VVH}
\left( \frac{1}{\lambda}\,-\,\lambda\,\hat{H}_0 \right)^2 \ =\
\frac{1}{\lambda^2} \,-\,\hat{V}^2_j  \,.
\end{equation}
This relation can be rewritten in two equivalent forms
\begin{equation} \label{V^2=H}
\hat{V}^2_j \ =\ 2\hat{H}_0\,-\,\lambda ^2\, \hat{H}^2_0 \,,
\end{equation}
\begin{equation}\label{H=V^2}
\hat{H}_0\, =\, \frac{1}{\lambda ^2} \left(1\,-\,\sqrt{1-\lambda^2\, \hat{V}^2}\,\right) \,.
\end{equation}

1) The first equation \eqref{V^2=H} implies that the kinetic energy of a particle may not be infinitely large, instead, it has a natural cut-off at $E_1\, =\, 2/\lambda^2$, since the l.h.s. of (\ref{V^2=H}) is determined by a positive operator.

2) Alternatively, from \eqref{H=V^2} it follows that with the kinetic energy $E_0  = \frac{1}{\lambda^2}$ one achieves the maximal value of $\hat{V}^2_j = \frac{1}{\lambda^2}$, since for higher velocities $H_0$ would not be hermitian.

The operators $\hat{L}_{ab}$ and $\hat{V}_a$, $a,b\,=\,1,\,..., \,4$ are generators of the group $E(4) = SO(4)\triangleright T(4)$ (= semi-direct product of orthogonal group $SO(4)$ and 4D translations $T(4)$). Adding now to \eqref{V^2=H} the square of the fourth component of the velocity operator \eqref{V4} we obtain the quadratic Casimir operator of the  $E(4)$ group:
\begin{equation} \label{casim2}
\hat{C}_2\ =\ \hat{V}^2_a \ =\ \hat{V}^2_j \,+\, \hat{V}_4^2 \ =\
2\,\hat{H}_0\,-\,\lambda ^2\, \hat{H}^2_0 \,+\, \left( \frac{1}{\lambda}\,-\,\lambda\, \hat{H}_0\right)^2 \ =\ \frac{1}{\lambda^2} \ .
\end{equation}
The second quartic $ISO(4)$ Casimir operator is given as a square of the $E(4)$ Pauli-Lubanski vector
\begin{equation} \label{PL}
\hat{\Lambda}_d\ =\ \frac{1}{2}\,\varepsilon_{abcd}\,\hat{L}_{ab}\,\hat{V}_d \ \ \ \Longleftrightarrow \ \ \ \hat{\Lambda}_i\ =\ \hat{V_4}\,\hat{L}_i \,+\, \varepsilon_{ijk}\,\hat{V}_j\,\hat{L}_{k4}, \ \hat{\Lambda}_4\ =\ \hat{L}_j\,\hat{V}_j \,.
\end{equation}
Particularly, the action of the fourth component of the Pauli-Lubanski vector is evaluated in the Appendix. The result is
\begin{equation}
\hat{\Lambda}_4\,\psi\ =\ \hat{L}_j\,\hat{V}_j\,\psi\ =\ 0\,.
\end{equation}
It follows from the $SO(4)$ invariance  that all four components of Pauli-Lubanski vector vanish, and consequently the quartic Casimir operator vanishes too:
\begin{equation} \label{casim4}
\hat{\Lambda}_a\,=\,0,\ \ a\,=\,1,\,..., \,4,\ \ \ \Longrightarrow \ \ \ \hat{C}_4\ =\ \hat{\Lambda}^2_a \ =\ 0\,.
\end{equation}

Thus, the NC QM in question is specified by a scalar $ISO(4)$ representation specified by the values of Casimir operators: $\hat{C}_2\,=\, 1/\lambda^2$ and $\hat{C}_4\,=\,0$. In such representation the common eigenvalues of velocity operators form a 3-sphere $S^3_\nu$ with radius $\nu\,=\,\frac{1}{\lambda}$ for any central potential $\hat{U}(\hat{r})$.

\subsection*{The acceleration operator}

In Newtonian mechanics the time derivative of a particle momentum $\vec{p}$ is governed by the equation $ \dot{\vec{p}}\, =\, - \vec{\nabla} U(\vec{x})$. In standard QM this is replaced by the  Heisenberg equation for $\vec{p}$:
$$ \frac{d}{dt} \vec{p}\ =\ -i\,[\,\vec{p},\,H]\ =\  -\,\vec{\nabla}U(\vec{x}) \,.$$

One may ask how is this relation modified in NC QM. The result for $\hat{U}=U(\hat{r})$ is easily calculated using previous results: \eqref{VV} combined with \eqref{H=V^2} that tells us that $[\hat{V}_i, \hat{H}_0]=0$. Consequently, the time derivative of a particle momentum, or its velocity using the units $m=1$,  is given as
\begin{equation} \label{ehren}
\dot{\hat{V}}_i\  =\ -i\, [\hat{V}_i ,\, \hat{H}]\, =\, -i\, [ \hat{V}_i ,\, \hat{H}_0\, +\, \hat{U}(\hat{r})]\ =\ -i\, [ \hat{V}_i ,\,\hat{U}(\hat{r})]\,.
\end{equation}
The last commutator is calculated in Appendix. Due to the modified Leibniz rule \eqref{Leibnitz}, the acceleration is equal to
\begin{equation} \label{Acc}
\dot{\hat{V}}_i\ =\ -i\, (\hat{V}_i\hat{U}(\hat{r}))\, +\, \hat{U}_\lambda'(\hat{r}) \left(\frac{\lambda}{\hat{r}}\,\hat{L}_i \,+\, \lambda^2\, \hat{W}_i \right) \,+\, \frac{\lambda^2}{2}\, \hat{U}_\lambda'' \,(\hat{r})\hat{V}_i \,,
\end{equation}
where $\hat{U}_\lambda'(\hat{r})$ is defined in \eqref{f'} and
\begin{equation}\label{U''}
\hat{U}''_\lambda(\hat{r})\ =\ \frac{1}{\lambda^2}\,\left(\hat{U}(\hat{r}+\lambda)\, -\, 2\,\hat{U}(\hat{r})\, +\, \hat{U}(\hat{r}-\lambda) \right)\,,
\end{equation}
\begin{equation}
\hat{W}_i\, \psi\ =\, \frac{1}{2r}\,\sigma^i_{\alpha \beta}\, [a_\beta,\, [a^+_\alpha ,\, \psi]] \,.
\end{equation}
The NC acceleration operator contains new terms that vanish in the commutative limit $\lambda \rightarrow 0$. The vector $\hat{W}_i$ has not been introduced yet, but it turns out to be closely related to the kinetic part of the Laplace - Runge - Lenz vector, see \cite{LRL}.

Time derivative of velocity expectation is fully determined by expectation of this commutator (Ehrenfest theorem).

\section{Summary and conclusions}

In this paper, in the framework of  NC QM, we have analyzed properties of the velocity operator $\hat{V}_j \, =\,-i\,[\hat{X}_j,\,\hat{H}]$. In standard QM, using units $\hbar\,=\,m\,=\,1$, the velocity operator is proportional to the gradient operator: $V_j^{QM} \, =\,-i\,\partial_j$. In NC QM the corresponding formula for velocity operator  $V_j^{NC}$ is given in \eqref{V1}. We found that its action on simple functions, such as $x_i$ or $r$, is closely related to the commutative case, see \eqref{VX}, \eqref{Vr}. However, there is a crucial difference, as $V_j^{NC}$ does not satisfy ordinary Leibniz rule, but only modified Leibniz rule \eqref{Leibnitz} with a NC correction \eqref{kor}. With non-commuting coefficients at commutators (derivatives) this is inevitable, and it leads to various modifications of the standard results:

$\bullet $ Relation  $[\hat{V}_i, \hat{V}_j]\,=0\,$ is the same as in the ordinary case, but is true only for states $\psi$ with a classical analogue (they have the same number of c/a operators). Even thought this result seems trivial, $[\hat{V}_i, \hat{V}_j]\,\neq\,0$ for states with a different number of c/a operators. 
Other results, listed below, give non trivial corrections also for states with equal number of c/a operators (corresponding to $\psi = \psi (\vec{x})$).

$\bullet $ The commutation relation \eqref{UR} between the position and velocity operators replaces the usual r. h. s. of the commutator $-i\delta^{ij}$ by $-i\delta^{ij}(1-\lambda^2 H_0)$, providing two important energy scales $E_0=\frac{1}{\lambda^2}$ and $E_1 = \frac{2}{\lambda^2}$. They are nonperturbative, as  they go to infinity in the commutative limit $\lambda \to 0$.

$\bullet $ The relation between the velocity operator and the free Hamiltonian \eqref{VVH}, equivalently \eqref{V^2=H} or \eqref{H=V^2}, provides interpretation of the nonperturbative energy scales mentioned above: at energy $E_0=\frac{1}{\lambda^2}$ the velocity gets its maximal value (corresponding to $V^2=\frac{1}{\lambda^2}$), and  $E_1 = \frac{2}{\lambda^2}$ represents, for any central potential, a universal kinetic energy cut-off.

$\bullet $ The formula for the time-derivative of velocity \eqref{Acc} shows that not only the gradient of the potential energy defines the acceleration of the particle, but also additional terms  depending on the derivatives of the potential, that, however, vanish for $\lambda=0$.

$\bullet $ All these results sum up in canonical (general not depending on a particular form of central potential) commutation relations $E(4) = SO(4)\triangleright T(4)$ supplemented by quadratic relations:
\begin{eqnarray}
\nn [\hat{L}_{ab},\,\hat{L}_{cd}] &=& i\,(\delta_{ac}\,\hat{L}_{bd}\,-\, \delta_{bc}\,\hat{L}_{ad} \,-\, \delta_{ad}\,\hat{L}_{bd}\,+\, \delta_{bd}\,\hat{L}_{ac} )\,, \\
\label{E4} [\hat{L}_{ab},\,\hat{V}_c]  &=& i\,(\delta_{ac}\,\hat{V}_b\,-\, \delta_{bc}\,\hat{V}_a )\,,\ \ \ [\hat{V}_a,\,\hat{V}_b] \ =\ 0\,, \\
\label{CE4}
\hat{V}^2_a  &=& \frac{1}{\lambda^2}\,,\ \ \ \varepsilon_{abcd}\,\hat{L}_{ab}\,\hat{V}_d \ =\ 0\,.
\end{eqnarray}
The quadratic relations specify the unitary irreducible $E(4)$ representation that can be realized in the space of functions on a 3-sphere with radius $1/\lambda$: $\Psi(v)$, $v \,\in\,S^3_{1/\lambda}$. In the commutative limit it reduces the standard  3D momentum/velocity representation.

All these general results are fully consistent with explicit solutions of the NC Schr\"odinger equation found recently for Coulomb problem in the NC space in question, \cite{GP1}, \cite{GP2}, \cite{LRL}.

\section*{Appendix}

For the calculations it is useful to define two sets of auxiliary operators, which act as follows
\begin{eqnarray} \label{auxAB}
\hat{a}_\alpha \psi\ =\ a_\alpha \psi\,, & \ \ \hat{a}^+_\alpha \psi\ =\ a^+_\alpha \psi\,, \\ \nonumber
\hat{b}_\alpha \psi\ =\ \psi a_\alpha\,, & \ \ \hat{b}^+_\alpha \psi\ =\ \psi a^+_\alpha\,.
\end{eqnarray}
While a-operators obviously commute with b-operators, the non vanishing commutators of two a-operators and two b-operators are
\be\label{ab} [\hat{a}_\alpha , \hat{a}^+_\beta]\ =\ \delta_{\alpha \beta}\,,\ \ \ [\hat{b}_\alpha , \hat{b}^+_\beta]\ =\ -\,\delta_{\alpha \beta}\,. \ee

\subsubsection*{ A. Derivation of the velocity operator $\boldsymbol{\hat{V}_j}$}
Here we will derive the velocity operator in form \eqref{V2.0}. Note, that for states with equal number of c/a operators it is true that
\begin{equation} \label{Hzeta}
\hat{H}_0\ =\ \frac{1}{2\lambda \hat{r}} \left(\frac{2 \hat{r}}{\lambda} - \hat{a}^+_\alpha \hat{b}_\alpha -\hat{b}^+_\alpha \hat{a}_\alpha \right) \,.
\end{equation}
Inserting \eqref{Hzeta} into \eqref{V} we obtain (recall that $[x^i, r]=0$):
\begin{equation}
\hat{V}_i\ =\ - \frac{i}{2\lambda \hat{r}}\, \left[\hat{X}_i ,\, \frac{2 \hat{r}}{\lambda}\, -\, \hat{a}^+_\alpha \hat{b}_\alpha -\hat{b}^+_\alpha \hat{a}_\alpha \right]\ =\ \frac{i}{2\lambda \hat{r}}\, \left[\hat{X}_i ,\, \hat{a}^+_\alpha \hat{b}_\alpha +\hat{b}^+_\alpha \hat{a}_\alpha \right] \,.
\end{equation}
The coordinate operator in terms of a- and b-operators can be written as
$$\hat{X}_i\, =\,\frac{\lambda}{2}\, \sigma^j_{\alpha \beta}\, (\hat{a}^+_\alpha \hat{a}_\beta\, +\, \hat{b}^+_\alpha \hat{b}_\beta)\,. $$
Using \eqref{ab} we obtain the final result
\begin{equation}
\hat{V}_i\ =\ \frac{i}{2r}\,  \sigma^i_{\alpha \beta}\,\hat{w}_{\alpha \beta}\,,\ \ \ \hat{w}_{\alpha \beta}\ =\ (\hat{a}^+_\alpha \hat{b}_\beta - \hat{a}_\beta \hat{b}^+_\alpha) \,.
\end{equation}

\subsubsection*{ B. Evaluation of the correction terms from the uncertainty relation}

We need to evaluate two correction terms:
\begin{eqnarray}
{\cal K}_i (x_j,\psi) &=& -\frac{i}{2r}\,\sigma ^i _{\alpha \beta} \left([a^+_\alpha ,\, x_j]\,[a_\beta ,\, \psi]\,-\, [a_\beta ,\, x_j]\,[a^+_\alpha ,\, \psi] \right)\,,\\ \nonumber
{\cal K}_i (\psi,x_j) &=& -\frac{i}{2r}\,\sigma ^i _{\alpha \beta} \left([a^+_\alpha ,\,\psi]\,[a_\beta ,\, x_j]\,-\, [a_\beta ,\, \psi]\,[a^+_\alpha ,\,  x_j] \right)\,.
\end{eqnarray}
Using \eqref{bosonic}, \eqref{X} and properties of Pauli matrices we get
\begin{eqnarray}
{\cal K}_i (x_j,\psi) &=& -\frac{i}{2r}\sigma ^i _{\alpha \beta} (\overbrace{[a^+_\alpha , x_j]}^{-\lambda \sigma^j_{\gamma \alpha} a^+_\gamma}[a_\beta , \psi]- \overbrace{[a_\beta , x_j]}^{\lambda\sigma^j_{\beta \delta} a_\delta}[a^+_\alpha , \psi])) \\ \nonumber
&=&\frac{i \lambda}{2 r}(\underbrace{\sigma^i_{\alpha \beta}\sigma^j_{\gamma \alpha}}_{\delta^{ij}\delta_{\gamma \beta} + i \varepsilon^{jik}\sigma^k_{\gamma \beta}}  a^+_\gamma (a_\beta \psi - \psi a_\beta) +\underbrace{\sigma^i_{\alpha \beta} \sigma^j_{\beta \delta}}_{\delta^{ij}\delta_{\alpha \delta} + i \varepsilon^{ijk}\sigma^k_{\alpha \delta}} a_\delta(a^+_\alpha \psi - \psi a^+_\alpha))
\end{eqnarray}
and similarly for ${\cal K}_i (\psi,x_j)$
\be
{\cal K}_i (\psi,x_j)\ =\ -\frac{i\lambda}{2r}((a^+_\alpha \psi - \psi a^+_\alpha) \sigma^i_{\alpha \beta} \sigma^j _{\beta \gamma} a_\gamma + (a_\beta \psi - \psi a_\beta) \sigma^i_{\alpha \beta} \sigma^j_{\gamma \alpha} a^+_\gamma)
\ee
Combing those two results we have
\begin{eqnarray}
{\cal K}_i (x_j,\psi) + {\cal K}_i (\psi,x_j) &=&\delta^{ij}\,((a^+_\alpha a_\alpha \psi - a^+_\alpha \psi a_\alpha + a_\alpha a^+_\alpha \psi - a_\alpha \psi a^+_\alpha)\\ \nonumber
&+&(-a_\alpha \psi a^+_\alpha + \psi a^+_\alpha a_\alpha - a^+_\alpha \psi a_\alpha + \psi a^+_\alpha a_\alpha)) \\ \nonumber
&+&i\varepsilon^{ijk} \sigma^k_{\alpha \beta}\,((-a^+_\alpha a_\beta \psi+ a^+_\alpha \psi a_\beta + a_\beta a^+_\alpha \psi - a_\beta \psi a^+_\alpha )\\ \nonumber
 &-&(a^+_\alpha \psi a_\beta - \psi a^+_\alpha a_\beta - a_\beta \psi a^+_\alpha + \psi a_\beta a^+_\alpha)) \,.
\end{eqnarray}
Using \eqref{bosonic} and the fact, that $Tr\ \sigma^k =0$, it is evident that terms in third and fourth line add up to zero. On the other hand, considering the equation \eqref{Hzeta} it is obvious that
\begin{equation}
\frac{1}{2}\,\left({\cal K}_i (x_j,\psi)\, +\, {\cal K}_i (\psi,x_j) \right)\ =\  i\,\delta^{ij}\,\lambda^2\, H_0\, \psi \,.
\end{equation}

\subsubsection*{C. The commutator $\boldsymbol{[\hat{V}_i, \hat{V}_j]}$}

As this calculation is quite complex, we will just outline it here. $[\hat{V}_i, \hat{V}_j]$ is antisymmetric, so we can calculate $\varepsilon^{ijk}[\hat{V}_i, \hat{V}_j]$ instead. In fact we will do so, because of the vector Fierz identity $\varepsilon ^{ijk} \sigma ^ i _{\alpha \beta} \sigma ^ j _{\gamma \delta} = i (\sigma ^k _{\alpha \delta} \delta_{\gamma \beta} - \sigma ^k _{\gamma \beta} \delta _{\alpha \delta})$, which we want to use. Using the notation \eqref{V2.0} we have
\begin{eqnarray}
\varepsilon^{ijk}[\hat{V}_i,\, \hat{V}_j] &=& \left(\frac{i}{2}\right)^2\,\varepsilon^{ijk}\, \sigma^i_{\alpha \beta}\, \sigma^j _{\gamma \delta}\, \left[\frac{1}{\hat{r}} \hat{w}_{\alpha \beta} ,\, \frac{1}{\hat{r}}\hat{w}_{\gamma \delta}\right] \\ \nonumber
&=& \left(\frac{i}{2}\right)^2 \,i \left(\sigma ^k _{\alpha \delta}\, \delta_{\gamma \beta}\, -\, \sigma ^k _{\gamma \beta}\, \delta _{\alpha \delta}\right) \\ \nonumber
&\times& \left(\frac{1}{\hat{r}^2}\,[\hat{w}_{\alpha \beta} , \hat{w}_{\gamma \delta}]\,+\, \frac{1}{\hat{r}}\,[\hat{w}_{\alpha \beta},\, \frac{1}{\hat{r}}]\,\frac{1}{\hat{r}}\,+\,\frac{1}{\hat{r}}\,[\frac{1}{\hat{r}},\,\hat{w}_{\gamma \delta}]\,\hat{w}_{\alpha \beta}\right) \,.
\end{eqnarray}
The contribution from the first term in the third line is easily evaluated using \eqref{V2.0} and \eqref{auxAB}, yielding
\begin{equation} \label{VVv1}
\varepsilon^{ijk}\, \sigma^i_{\alpha \beta}\, \sigma^j_{\gamma \delta}\, \frac{1}{\hat{r}^2}\,[\hat{w}_{\alpha \beta} ,\, \hat{w}_{\gamma \delta}]\ =\ \frac{8i}{\hat{r}^2}\, \hat{L}_k \,.
\end{equation}
The contribution from the other two terms in the third line is a bit more demanding. As we stated earlier, the functions of $\hat{r}$ are defined by the Taylor expansion, therefore we need to evaluate the commutators of c/a operators with powers of $r$. This can be done using $[a_\alpha , r] =\lambda a_\alpha$, $[a^+_\alpha , r] = -\lambda a^+_\alpha$ in a following way:
\begin{equation} \label{arN}
a_\alpha\, r^N \ =\ a_\alpha\, \underbrace{r...r}_N \ =\ (r+\lambda)\ a_\alpha\, \underbrace{r...r}_{N-1} \ =\ \cdots\ =\ (r+\lambda)^N \, a_\alpha \,,
\end{equation}
\begin{equation*}
a^+_\alpha\, r^N \ =\ (r-\lambda)^N \,a^+_\alpha \,.
\end{equation*}
Using those relations we get
\begin{equation} \label{VVv2}
\varepsilon^{ijk}\,\sigma^i_{\alpha \beta}\, \sigma^j_{\gamma \delta} \left(\frac{1}{\hat{r}}\,[\hat{w}_{\alpha \beta},\, \frac{1}{\hat{r}}]\,\frac{1}{\hat{r}}\,+\,\frac{1}{\hat{r}}\,[\frac{1}{\hat{r}},\,\hat{w}_{\gamma \delta}]\,\hat{w}_{\alpha \beta} \right)\ =\ \cdots\ =\ -\,\frac{8i}{\hat{r}^2}\,\hat{L}_k \,.
\end{equation}
Adding together results \eqref{VVv1} and \eqref{VVv2} gives $\varepsilon^{ijk}\,[\hat{V}_i,\, \hat{V}_j]\,=\,0$, and therefore $[\hat{V}_i,\, \hat{V}_j]\,=\,0$.

\subsubsection*{ D. Relation between the velocity operator and the free Hamiltonian}

Derivation of this relation requires a number of trivial steps and one tricky step. We will omit the trivial ones here, showing only the important steps. Writing down the velocity operator(s) using \eqref{V2.0} we have
\begin{eqnarray} \label{Vdva}
\hat{V}^2  &=&  \frac{i}{2\hat{r}}\,\sigma^i_{\alpha \beta} \left(\hat{a}^+_\alpha\, \hat{b}_\beta\, -\, \hat{a}_\beta\, \hat{b}^+_\alpha \right)\,\sigma^j_{\gamma \delta}\,\frac{i}{2\hat{r}}\left(\hat{a}^+_\gamma\, \hat{b}_\delta\, -\, \hat{a}_\delta \hat{b}^+_\gamma \right) \\ \nonumber
&=& \cdots\ \ \ \mbox{a lot of trivial modifications}\ \ \ \cdots \\ \nonumber
&=& \frac{1}{\lambda^2}\, -\, \frac{1}{4\hat{r}(\hat{r}-\lambda)} \left((\hat{a}^+\,\hat{b})^2\,+\,(\hat{a}^+\,\hat{b})\,(\hat{a}\,\hat{b}^+))\right) \\ \nonumber
&-& \frac{1}{4\hat{r}(\hat{r}+\lambda)} \left((\hat{a}\,\hat{b}^+)^2\,+\,(\hat{a}\hat{b}^+)\,(\hat{a}^+\,\hat{b}) \right) \,,
\end{eqnarray}
where $(\hat{a}^+\hat{b})=\hat{a}^+_\alpha \hat{b}_\alpha$ and similarly for other combinations. From \eqref{Hzeta} it is evident that
\begin{equation} \label{Hdva}
\hat{H}_0\, -\, \frac{1}{\lambda^2}\ =\ -\, \frac{1}{2\lambda \hat{r}} \left((\hat{a}^+\,\hat{b})\,+\,(\hat{b}^+\,\hat{a})\right) \,.
\end{equation}
While in \eqref{Vdva} there are four c/a operators in each term, there are only two of them in \eqref{Hdva}, so we need to square  \eqref{Hdva}:
\begin{eqnarray}   \label{V^2b}
\left(\frac{1}{\lambda ^2}\, -\, \hat{H}_0 \right)^2 &=& \frac{1}{\lambda^2 \hat{r}}\left( (\hat{a}^+\,\hat{b})\, +\, (\hat{b}^+\,\hat{a})\right)\,\frac{1}{\hat{r}}\left((\hat{a}^+\,\hat{b})\, +\, (\hat{b}^+\,\hat{a})\right) \\ \nonumber
&=& \frac{1}{4\lambda^2\hat{r}(\hat{r}-\lambda)} \left((\hat{a}^+\,\hat{b})^2\,+\,(\hat{a}^+\,\hat{b})\,(\hat{a}\,\hat{b}^+)\right) \\ \nonumber
&+& \frac{1}{4\lambda^2\hat{r} (\hat{r}+\lambda)}\left((\hat{a}\,\hat{b}^+)^2\, +\, (\hat{a}\,\hat{b}^+)\,(\hat{a}^+\,\hat{b}) \right)\,.
\end{eqnarray}
In the last step we have used \eqref{arN}. Now comparing this result with \eqref{Vdva} it is evident that
\begin{equation}
\left(\frac{1}{\lambda^2}\,-\,\hat{H}_0 \right)^2 = \frac{1}{\lambda ^2} \left(\frac{1}{\lambda^2}\,-\,\hat{V}^2 \right) .
\end{equation}

\subsubsection*{E. The acceleration operator}

The idea of this calculation is to evaluate the correction term \eqref{kor} for \eqref{ehren}, which with use of \eqref{arN} turns out to be
\begin{equation} \label{KU}
{\cal K}_i ( U(r), \psi)\ =\ -\frac{i}{2r}\,\sigma^i_{\alpha \beta} \left([a^+_\alpha ,\,U(r)]\,[a_\beta,\,\psi]\ -\ [a_\beta,\, U(r)]\,[a^+_\alpha ,\, \psi] \right)
\end{equation}
\begin{equation*}
=\ -\frac{i}{2r}\sigma^i_{\alpha \beta} \ ((U(r-\lambda)-U(r))\underbrace{a^+_\alpha[a_\beta,\psi]}_{\hat{A}_{\alpha \beta} \psi}\ -\ (U(r+\lambda)-U(r))\underbrace{a_\beta[a^+_\alpha , \psi]}_{\hat{B}_{\alpha \beta} \psi} \ )\,.
\end{equation*}
Now we need to evaluate the underbraced terms. As one can easily check, they are equal to
\begin{eqnarray} \label{AB}
\hat{A}_{\alpha \beta}&=&\frac{1}{2} \left(\hat{{\cal W}}_{\alpha \beta} + {\cal L}_{\alpha \beta} - \hat{w}_{\alpha \beta}\right)\,, \\ \nonumber
\hat{B}_{\alpha \beta}&=&\frac{1}{2}\left(\hat{{\cal W}}_{\alpha \beta} + {\cal L}_{\alpha \beta} + \hat{w}_{\alpha \beta}\right)\,,
\end{eqnarray}
where
$$\hat{{\cal W}}_{\alpha \beta}\ =\ \hat{a}^+_\alpha \,\hat{a}_\beta \,+\, \hat{b}^+_\alpha \,\hat{b}_\alpha \,-\, \hat{a}^+_\alpha\, \hat{b}_\beta\, -\ \hat{a}_\beta\, \hat{b}^+_\alpha\,,\ \ \ \frac{1}{2}\,\sigma^i_{\alpha \beta}\, {\cal L}_{\alpha \beta}\ =\ \hat{L}^i $$
and $\hat{w}_{\alpha \beta}$ is defined in \eqref{V2.0}. Such decomposition might seem a bit artificial, but will become more transparent in \cite{LRL}. Inserting \eqref{AB} into \eqref{KU} we obtain the result
\begin{eqnarray}
-i\,{\cal K}_i(U(r) , \psi) &=& \left(\hat{U}_\lambda'(\hat{r}) \,(\frac{\lambda}{\hat{r}}\,\hat{L}_i\, +\, \lambda^2 \hat{W}_i)\,+\, \frac{\lambda^2}{2} \,\hat{U}_\lambda''(\hat{r})\hat{V}_i \right) \psi\,,\\ \nonumber \hat{W}_i &=& \frac{1}{2}\,\sigma^i_{\alpha \beta}\, {\cal W}_{\alpha \beta} \,.
\end{eqnarray}
The second order difference $\hat{U}''$ is specified in \eqref{U''}.

\nocite{*}
\bibliography{aipsamp}%

\begin{thebibliography}{99}

\bibitem{Snyder} H. S. Snyder, Quantized space-time, Phys. Rev. 71 (1947) 38.
%
\bibitem{Yang} C. N. Yang, On quantized space-time Phys. Rev. 72 (1947) 874.
%
\bibitem{Wheeler} J. A. Wheeler,  {\it Geometrodynamics}, Academic Press, New York (1962).
%


\bibitem{Con} A. Connes, Publ. IHES {\bf 62} (1986) 257;
A. Connes, {\it Noncommutative Geometry} (Academic Press, London, 1994).
%
\bibitem{Mad1} M. Dubois-Violete, {\it C. R. Acad. Sci. Paris} {\bf 307}
(1988) 403; M. Dubois-Violete, R. Kerner and J. Madore, J. Math. Phys.
{\bf 31} (1990) 316.
%
\bibitem{DFR} S. Doplicher, K. Fredenhagen, J. F. Roberts, Comm. Math. Phys. {\bf 172} (1995) 187.
%
\bibitem{string} M. M. Sheikh-Jabbari, Phuys. Lett {\bf B425} (1998) 48; V. Schomerus, JHEP {\bf  9906} (1999) 030; N. Seiberg and E. Witten JHEP {\bf  9909} 97.
%
\bibitem{AB} M. Chaichian, A. Demichev, P. Pre\v{s}najder, M. M.  Sheikh-Jabbari, A. Tureanu,
Nucl.Phys. {\bf B 611 }(2001) 383; M. Chaichian, A. Demichev, P. Pre\v{s}najder, M.M.
Sheikh-Jabbari and A. Tureanu, Phys. Lett. {\bf B527} (2002) 149;
H. Falomir, J. Gamboa, M. Loewe and J. C. Rojas, Phys. Rev. {\bf
D66} (2002) 045018; M. Chaichian, Miklos Langvik, Shin Sasaki and
Anca Tureanu, Phys. Lett. {\bf B666} (2008) 199.
%
\bibitem{CP} M. Chaichian, M.M. Sheikh-Jabbari and A. Tureanu,
Phys. Rev. Lett. {\bf 86} (2001) 2761; M. Chaichian, M. M.
Sheikh-Jabbari and A. Tureanu, Eur. Phys. J. {\bf C36} (2004) 251;
T. C. Adorno, M. C. Baldiotti, M. Chaichian, D. M. Gitman and A.
Tureanu, Phys. Lett. {\bf B682} (2009) 235.
%
\bibitem{Scholtz} F. G. Scholtz, B. Chakraborty, J. Goaverts, S. Vaidya,
J. Phys. A: Math. Theor. {\bf A40} (2007) 14581; J. D. Thom, F. G.
Scholtz, J. Phys. A: Math. Theor. {\bf A42} (2009) 445301.
%
\bibitem{GP1} V. G\'alikov\'a, P. Pre\v snajder,
J. Phys.: Conf. Ser. {\bf 343} (2012 ) 012096.
%
\bibitem{GP2} V. G\'alikov\'a, P. Pre\v snajder, J. Math. Phys. {\bf 54} (2013) 052102.
%
\bibitem{LRL} V. G\'alikov\'a, S. Kov\'a\v{c}ik, P. Pre\v{s}najder, {\it Laplace-Runge-Lenz vector for Coulomb problem in NC quantum mechanics} - in preparation.
%
\end{thebibliography}

\end{document}